\RequirePackage{amsfonts,amssymb,amsmath}
\documentclass{article}
\usepackage[utf8]{inputenc}
\usepackage[natbib,pdf,cref,theorems,ruby]{nikis} 
\usepackage{fullpage}
\usepackage{anyfontsize}
\usepackage[group-separator={,}]{siunitx}
\usepackage{wrapfig}
\usepackage{floatrow}
\usepackage{booktabs}
\usepackage[caption=false]{subfig}

\newcommand*\xor{\ensuremath{\mathbin{\oplus}}}
\newcommand{\lcg}[1]{\ensuremath{\instancename{lcg}_{#1}}}
\newcommand{\sxor}[1]{\instancename{sxor}_{#1}}
\newcommand{\id}{\instancename{id}}

\usepackage{forest}
\newcommand{\block}[1]{\noindent\textsl{#1. }}

\usepackage{bibunits}
\defaultbibliography{ref}
\defaultbibliographystyle{abbrvnat}

\newcommand{\vRolling}[1][]{\instancename{rolling#1}}
\newcommand{\vCompact}{\instancename{cht}}
\newcommand{\vHash}[1][] {\instancename{hash#1}}
\newcommand{\vBonsai} {\instancename{bonsai}}
\newcommand{\vJudy}   {\instancename{judy}}
\newcommand{\vCedar}  {\instancename{cedar}}
\newcommand{\vTernary}{\instancename{ternary}}
\newcommand{\vBinary} {\instancename{binary}}

\newcommand{\hashfunction}[1]{{\usefont{T1}{ppl}{m}{sl}#1}}

\newcommand{\vFermat} {\hashfunction{fermat}}
\newcommand{\vIDThree}{\hashfunction{ID37}}

\newcommand{\JO}[2][]{#1}

\title{Practical Evaluation of Lempel-Ziv-78 and Lempel-Ziv-Welch Tries}
\author{%
  Johannes Fischer%
  \and
  Dominik K\"{o}ppl%
}

\date{Department of Computer Science, TU Dortmund, Germany}
\begin{document}
\begin{bibunit}

\maketitle
\begin{abstract}
We present the first thorough practical study of the Lempel-Ziv-78 and the Lempel-Ziv-Welch computation
based on trie data structures. 
With a careful selection of trie representations we can beat well-tuned 
popular trie data structures like Judy, m-Bonsai or Cedar. 
\end{abstract}
\pagestyle{plain}

\section{Introduction}

The LZ78-compression scheme \cite{ziv78compression} is an old compression scheme that is still in use today, e.g., in the Unix \texttt{compress} utility, in the GIF-standard, in string dictionaries \cite{arz14lz}, or in text indexes \cite{ArroyueloN11}. Its biggest advantage over LZ77 \cite{ziv77universal} is that LZ78 allows for an easy construction \emph{within compressed space} and in \emph{near-linear time}, which is (to date) not possible for LZ77. Still, although LZ77 often achieves marginally better compression rates, the output of LZ78 is usually small enough to be used in practice, e.g.\ in the scenarios mentioned above \cite{belazzougui16range,ArroyueloN11}.

While the construction of LZ77 is well studied both in theory \cite[e.g.]{belazzougui16range,lzciss} and in practice \cite[e.g.]{kaerkkaeinen16lazy,kaerkkaeinen13lightweight}, only recent interest in LZ78 can be observed: just in 2015 \citet{nakashima15constructing} gave the first (theoretical) linear time algorithm for LZ78. On the practical side, we are not aware of any systematic study.

We present the first thorough study of LZ78-construction algorithms. Although we do not present any new theoretical results, this paper shows that if one is careful with the choices of tries, hash functions, and the handling of dynamic arrays, one can beat well-tuned out-of-the-box trie data structures like Judy\footnote{\url{http://judy.sourceforge.net}}, m-Bonsai~\cite{mBonsai}, or the Cedar-trie~\cite{cedarTrie}.

\block{Related Work}
An LZ78 factorization of size~$z$ can be stored in two arrays with $z \lg \sigma$ and $z \lg z$ bits to represent the character and the referred index, respectively, of each factor.
This space bound has not yet been achieved by any efficient trie data structure.
Closest to this bound is the approach of
\citet[Lemma~8]{ArroyueloN11}, taking $2z \lg z + z \lg \sigma + {\Oh{z}}$ bits and {\Oh{n(\lg\sigma + \lg\lg n)}} time for the LZ78 factorization.
Allowing \Oh{z \lg z} bits, {\Oh{n+z\frac{\lg^2\lg\sigma}{\lg\lg\lg\sigma}}} time is possible \cite{fischer15alphabet}.
Another option is the dynamic trie of \citet{Jansson2015LDT} using
$\Oh{n(\lg \sigma +  \lg \lg_{\sigma} n) / \lg_{\sigma} n}$ bits of working space and
$\Oh{n\lg^2 \lg n / \left(\lg_{\sigma} n \lg \lg \lg n\right)}$ time.
All these tries are favorable for small alphabet sizes (achieving linear or sub-linear time when $\lg \sigma = \oh{{\lg n \lg \lg \lg n}/{\lg^2 \lg n}}$).
If the alphabet size~$\sigma$ becomes large, the upper bounds on the time get unattractive.
Up to $\lg \sigma = \oh{\lg n}$, we can use a linear time solution taking  ${\Oh{n \lg \sigma}}$ bits of space~\cite{lzcics,MunroNN17}.
Finally, for large~$\sigma$, there is a linear time approach taking $(1+\epsilon) n \lg n + {\Oh{n}}$ bits of space~\cite{lzciss}.
Further \emph{practical} trie implementations are mentioned in Section \ref{sect:experiments}.

\section{Preliminaries}
Let $T$ be a text of length $n$ over an alphabet $\Sigma=\{1,.\dots,\sigma\}$ with $\abs{\Sigma} \le n^{\Oh{1}}$.
Given $X,Y,Z \in \Sigma^*$ with $T = XYZ$, 
then $X$, $Y$ and $Z$ are called a \intWort{prefix}, \intWort{substring} and \intWort{suffix} of $T$, respectively.
We call $T[i..]$ the $i$-th suffix of $T$, and denote a substring $T[i] \cdots T[j]$ with $T[i..j]$.
A \intWort{factorization} of $T$ of size~$z$ partitions $T$ into $z$~substrings (\intWort{factors}) $F_1 \cdots F_z = T$.
In this article, we are interested in the LZ78 and LZW factorization.
If we stipulate that $F_0$ and $F_{z+1}[1]$ are the empty string, we get:

A factorization $F_1\cdots F_z = T$ is called the \intWort{LZ78 factorization}~\cite{ziv78compression} of $T$ iff 
$F_x=F_y c$ with $F_y = \argmax_{S \in \menge{F_{y'} : 0 \le y' < x} } \abs{S}$ and $c\in\Sigma$
for all $1 \le x \le z$;
we say that $y$ is the \intWort{referred index} of the factor~$F_x$.

A factorization $F_1\cdots F_z = T$ is called the \intWort{LZW factorization}~\cite{Welch84} of $T$ iff 
$F_x=F_y F_{y+1}[1]$ with $F_y = \argmax_{S \in \menge{F_{y'} : 1 \le y' < x}} \abs{S}$, or $F_x = c \in \Sigma$ if no such $F_y$ exists,
for all $1 \le x < z$.
If~$F_x = F_y F_{y+1}[1]$ for a~$y$ with $1 \le y < x$, we call $y$ the \intWort{referred index} of the factor~$F_x$.
Otherwise, $F_x=c$ for a $c\in\Sigma$; we set its referred index to~$-c < 0$.

The factors can be represented in a trie, the so-called \intWort{LZ trie}.
Each factor~$F_x$ (except the last factor in LZW) is represented by a trie node $v$ labeled with~$x$ ($1\le x \le z$)
such that
the parent $u$ of $v$ is labeled with $y$ if $y$ is the referred index of $F_x$.
The edge~$(u,v)$ is then labeled with the last character of the factor~$F_x$ (or the first character of $F_{x+1}$ for LZW).

\block{Output}
We transform the list of factors to a list of integer values as follows:
We linearly process each factor~$F_x$ for $1 \le x \le z$.
If $F_x$'s referred index is not positive, $F_x$ is equal to a character~$c$ that is output (we output $-c$ in case of LZW). 
A factor $F_x$ with a referred index~$y > 0$ is processed as follows:
\begin{description}
	\item[LZ78:] If $F_x = F_y c$ for a $c \in \sigma$, we output the tuple~$(y,c)$.
	\item [LZW:] If $F_x = F_y F_{y+1}[1]$ (or $F_x = F_y$ for $x=z$), we output $y+\sigma$.
\end{description}

\block{Algorithm}
The folklore algorithm computing LZ78 and LZW uses a dynamic LZ trie that grows linearly in the number of processed factors.
The dynamic LZ trie supports the creation of a node, the navigation from a node to one of its children, and the access to the labels.

\begin{figure}[t]
	\floatbox[{\capbeside\thisfloatsetup{capbesideposition={right,center},capbesidewidth=6.75cm}}]{figure}[\FBwidth]%
	{%
	\caption{LZ78 trie and LZW trie. 
	Given the text $T = \texttt{aaababaaaba}$,
	LZ78 factorizes~$T$ into
		\RubyReset{}%
		\RubyCount{a}$\mid$%
		\RubyCount{aa}$\mid$%
		\RubyCount{b}$\mid$%
		\RubyCount{ab}$\mid$%
		\RubyCount{aaa}$\mid$%
		\RubyCount{ba}%
		,
		where the vertical bars separate the factors.
		The LZ78 factorization is output as:
		\texttt{a}$\mid$%
\texttt{(1,a)}$\mid$%
\texttt{b}$\mid$%
\texttt{(1,b)}$\mid$%
\texttt{(2,a)}$\mid$%
\texttt{(3,a)}.
		This output is represented by the left trie~(a).
		The LZW factorization of the same text is
		\RubyReset{}%
		\RubyCount{a}$\mid$%
		\RubyCount{aa}$\mid$%
		\RubyCount{b}$\mid$%
		\RubyCount{a}$\mid$%
		\RubyCount{ba}$\mid$%
		\RubyCount{aab}$\mid$%
		\RubyCount{a}.
		We output it as
\texttt{-1}$\mid$%
\texttt{1}$\mid$%
\texttt{-2}$\mid$%
\texttt{-1}$\mid$%
\texttt{3}$\mid$%
\texttt{2}$\mid$%
\texttt{-1}.
		This output induces the right trie~(b).
	}
	\label{figParsingTries}
}{%
	\centering{%
		\subfloat[LZ78-Trie\label{figParsingTries78}]{%
\tikzstyle{labelnode}=[inner sep=0pt,minimum size=8pt,sloped,above,midway,draw=none,font=\ttfamily]
\scalebox{0.7}{%
\begin{forest}
	for tree={circle,draw, l sep=20pt}
	[0	
		[1, edge label={node[labelnode] {a}} 
		[2, edge label={node[labelnode] {a}} 
		[5, edge label={node[labelnode] {a}} 
		]
	]
	[4, edge label={node[labelnode] {b}}]]
			[3, edge label={node[labelnode] {b}} 
			[6, edge label={node[labelnode] {a}} 
			]]]
\end{forest}
}%
}%
	\hspace{2em}
	\subfloat[LZW-Trie\label{figParsingTriesW}]{%
\tikzstyle{labelnode}=[inner sep=0pt,minimum size=8pt,sloped,above,midway,draw=none,font=\ttfamily]
\scalebox{0.7}{%
\begin{forest}
for tree={circle,draw, l sep=10pt}
[0	
	[-1, edge label={node[labelnode] {a}} 
	[1, edge label={node[labelnode] {a}} 
	[2, edge label={node[labelnode] {b}} 
	[6, edge label={node[labelnode] {a}} 
	]]
]
[4, edge label={node[labelnode] {b}}]]
		[-2, edge label={node[labelnode] {b}} 
		[3, edge label={node[labelnode] {a}} 
		[5, edge label={node[labelnode] {a}} 
		]]]]
\end{forest}
}%
}%
	}%
}%
\end{figure}
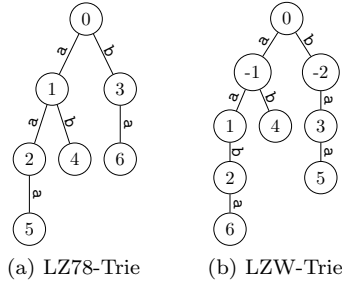

Given that $z$ is the number of LZ78 or LZW factors,
the algorithm performs $z$ searches of a prefix of a given suffix of the text.
It inserts $z$ times a new leaf in the LZ trie.
It takes $n$ times an edge from a node to one of its children.

\section{LZ-Trie representations}
In this section, we show five representations, each providing different trade-offs for computation speed and memory consumption.
All representations have in common that they work with dynamic arrays.

\block{Resize Hints}
The usual strategy for dynamic arrays is to double the size of an array when it gets full.
To reduce the memory consumption, a hint on how large the number of factors~$z$ might get is advantageous to know for a dynamic LZ trie data structure.
We provide such a hint based on the following lemma:
\begin{lemma}[{\cite{BannaiIT12,ziv78compression}}]\label{lemLZbounds}
	The number of LZ78 factors~$z$ is in the range 
	$\sqrt{2n+1/4} - 1/2 \le z \le cn / \lg_\sigma n$, for a fixed constant~$c > 0$.
\end{lemma}
At the beginning of the factorization, we let a dynamic trie reserve so much space such that it can store at least $\sqrt{2n}$ elements without resizing.
On enlarging a dynamic trie, we usually double its size.
But if the number of remaining characters~$r$ to parse is below a certain threshold, we try to scale
the data structure up to a value for which we expect that all factors can be stored without resizing the data structure again.
Let $z'$ be the currently computed number of factors.
If $r < n/2$ we use $z' + 3r/\lg r$ as an estimate (the number~3 is chosen empirically), derived from $z-z' = \Oh{r / \lg_{\sigma} r}$ based on \Cref{lemLZbounds}, 
otherwise we use $z' + z' r / (n-r)$ derived from the expectation that the ratio between $z'$ and $n-r$ will be roughly the same as 
between $z$ and $n$ (interpolation).

\subsection{Deterministic LZ Tries}

\begin{wrapfigure}[8]{r}{7cm}
\vspace{-3em}
		\begin{tabular}{l*{6}{c}}
			\toprule
			index & 1 & 2 & 3 & 4 & 5 & 6 
			\\\midrule
			first child & 2 & 5 & 6 & \\
			next sibling & 3 & 4 \\
			character & \texttt{a} & \texttt{a} & \texttt{b} & \texttt{b} & \texttt{a} & \texttt{a} 
			\\\bottomrule
		\end{tabular}
	\caption{Array data structures of \vBinary{} built on the example given in \Cref{figParsingTries}}
	\label{figFCNS}
\end{wrapfigure}
We first recall two trie implementations using arrays to store the node labeled with~$x$ at position~$x$, for each~$x$ with $1 \le x \le z$.\\
\block{Binary Search Trie}
The first-child next-sibling representation \vBinary{} maintains its nodes in three arrays.
A node stores a character, a pointer to one of its children, and a pointer to one of its siblings.
We do not sort the nodes in the trie according to the character on their incoming edge, but store them in the order in which they are inserted. (We found this faster in our experiments.)
\vBinary{} takes $2z \lg z + z \lg \sigma$ bits when storing $z$ nodes.
See \Cref{figFCNS} for an example.

\block{Ternary Search Trie}
A node of the Ternary Search Trie~\cite{BentleyS97} \vTernary{} stores a character,
a pointer to one of its children, a pointer to one of its smaller siblings, and a pointer to one of its larger siblings.
Similar to \vBinary{}, we do not rearrange the nodes.
\vTernary{} takes $3z \lg z + z \lg \sigma$ bits when storing $z$ nodes.

\subsection{LZ Tries with Hashing}
We use a hash table~$H[0..M-1]$ for a natural number~$M$, and a hash function~$h$ to store key-value pairs.
We determine the position of a pair~$(k,v)$ in~$H$ by the \intWort{initial address} $h(k) \mod M$;
we handle collisions with linear probing.
We enlarge~$H$ when the maximum number of entries~$m := \alpha M$ is reached,
where~$\alpha$ is a real number with $0 < \alpha < 1$. %

A hash table can simulate a trie as follows:
Given a trie edge $(u,v)$ with label $c$, we use the unique key $c + \sigma\ell$ to store $v$, where $\ell$ is the label (factor index) of $u$ (the root is assigned the label~$0$).
This allows us to find and create nodes in the trie by simulating top-down-traversals. 
This trie implementation is called \vHash{} in the following.

\block{Table Size}
We choose the hash table size~$M$ to be a power of two.
Having $M = 2^k$ for $k \in \N$, 
we can compute the remainder of the division of a hash value by the hash table size with a bitwise-AND operation,
i.e.,  $h(x) \mod 2^k = h(x) \& (2^{k}-1)$, which is practically faster\footnote{\url{http://blog.teamleadnet.com/2012/07/faster-division-and-modulo-operation.html}}.

If the aforementioned resize hint suggests that the next power of two is sufficient for storing all factors, we set $\alpha = 0.95$ before enlarging the size (if necessary).
We also implemented a hash table variant that will change its size to fit the provided hint.
This variant then cannot use the fast bit mask to simulate the operation $\hspace{-0.5em}\mod M$.
Instead, it uses a practical alternative that scales the hash value by~$M$ and divides this value by the largest possible hash value%
\footnote{\href{http://www.idryman.org/blog/2017/05/03/writing-a-damn-fast-hash-table-with-tiny-memory-footprints/}{http://www.idryman.org/blog/2017/05/03/writing-a-damn-fast-hash-table-with-tiny-memory-footprints/}}, i.e., 
$M h(k) / (\max_{k'} h(k'))$.
We mark those hash table variants with a plus sign, e.g., \vHash[+] is the respective variant of \vHash{}.

\subsubsection{Compact Hashing}
In terms of memory, \vHash{} is at a disadvantage compared to \vBinary{}, because
the key-value pairs consist of two factor indices and a character; for an~$\alpha < 1$,
\vHash{} will always take more space than \vBinary{}.
To reduce the size of the stored keys, we introduce the representation \vCompact{} using compact hashing.

The idea of compact hashing~\cite{knuthIII,FeldmanL73} is to use a bijective hash function such that 
when storing a tuple with key~$k$ in~$H$, we only store the value and the quotient~$\gauss{h(k)/M}$ 
in the hash table. The original key of an entry of~$H$ can be restored 
by knowing the initial address~$h(k) \mod M$ and the stored quotient~$\gauss{h(k)/M}$. 
To address collisions and therefore the displacement of a stored entry due to linear probing, 
\citet{Cleary84} adds two bit vectors with which the initial address can be restored.

For the bijective hash function~$h$, we consider two classes:

\block{The class of linear congruential generators~(LCGs)}
The class of LCGs~\cite{CarterW79} contains all functions 
$\lcg{a,b,p} : [0..p-1] \rightarrow [0..p-1], x \mapsto (a x + b) \mod p$ with $p \in \N, 0 < a < p, 0 \le b < p$.
If $p$ and $a$ are relative prime, then there exists a unique inverse~$a^{-1} \in [1..p-1]$ of $a$ such that
$a a^{-1} \mod p = 1$.
Then $\lcg{a,b,p}^{-1} : y \mapsto (y - b)a^{-1} \mod p$ is the inverse of $\lcg{a,b,p}$.
If $p$ is prime, then $a^{-1} = a^{p-2} \mod p$ due to Fermat's little theorem.

\block{The class of xorshift functions}
The xorshift hash function class~\cite{Marsaglia03} contains functions that use shift- and exclusive or (xor) operations.
Let $\xor$ denote the binary xor-operator and $w$ the number of bits of the input integer.
For an integer $j < -\gauss{w/2}$ or $j > \gauss{w/2}$, the xorshift operation 
$\sxor{w,j} : [0..2^w-1] \rightarrow [0..2^w-1], x \mapsto \left( x \xor (\gauss{2^j x} \mod 2^w) \right) \mod 2^w$ is inverse to itself: $\sxor{w,j} \circ \sxor{w,j} = \id$.

It is possible to create a bijective function that is a concatenation of functions of both families\footnote{%
Popular hash functions like MurmurHash 3 (\url{https://github.com/aappleby/smhasher}) use a post-processing step
that applies multiple LCGs~$\lcg{a,0,2^{64}}$ with $a$ as a predefined odd constant, and some xorshift-operations.}.

A compact hash table can use less space than a traditional hash table if the size of the keys is large:
If the largest integer key is~$u$, then all keys can be stored in $\upgauss{\lg u}$ bits, 
whereas all quotients can be stored in $\upgauss{\lg (\max_u h(u)/M)}$ bits.
By choosing the hash function carefully, it is possible to store the quotients in a number of bits independent of the number of the keys.

\block{Enlarging the hash table}
On enlarging the hash table, we choose a new hash function, and rebuild the entire table with the new size and a newly chosen hash function.
We first choose a hash function~$h$ out of the aforementioned bijective hash classes and adjust $h$'s parameters
such that $h$ maps from $[0..m\sigma-1]$ to $[0..2m\sigma-1]$ ($m$ has already its new size).
This means that

\begin{itemize}
	\item
	we select a function~$\lcg{a,b,p}$ with a prime $m\sigma < p < 2m\sigma$ (such a prime exists~\cite{Tchebychev} and can be precomputed for all $M=2^k$, $1 \le k \le \lg n$)
	and $0 < a,b \le p$ randomly chosen, or that
	\item we select a function~$\sxor{w,j}$ with $\lg(m\sigma) \le w \le \lg(2m\sigma)$ and $j$ arbitrary.
\end{itemize}

The hash table always stores trie nodes with labels that are at most~$m$; 
this is an invariant due to the following fact: before inserting a node with label~$m+1$ we enlarge the hash table and hence update~$m$.
Therefore, the key of a node can be represented by a $\upgauss{\lg(m\sigma)}$-bit integer (we map the key to a single integer with $[0..m-1]\times[0..\sigma-1]\rightarrow[0..m\sigma-1], (y,c) \mapsto \sigma y + c$).
Since $h$ is a bijection, 
the function $[0..m\sigma-1] \rightarrow [0..M-1] \times [0..\gauss{(2m\sigma-1)/M}],
i \mapsto (h_1(i),h_2(i)) := ( h(i) \mod M, \gauss{ h(i) / M })
$
is bijective, too.
We use $h_1$ to find the locations of the entries in of our hash table~$H$.
When we want to store a node with label~$x$ and key~$y\sigma+c$ in the hash table, we put $x$ and $h_2(\sigma y + c)$ in an entry of the hash table (the entry is determined by $h_1$, the linear probing strategy, and a re-arrangement with the bit vectors).
In total, we need $M \left( \lg(2\alpha\sigma) + \lg{m} \right) + 2M$ bits.
Since $m \le 2z -1$ there is a power of two such that $M = 2^{\gauss{\lg (z/\alpha)}+1} \le (2z-1)/\alpha$.
On termination, the compact hash table takes at most 
$M(2+ \lg(2\alpha\sigma m) ) \le (2z-1)(3+ \lg(\alpha\sigma z) )/\alpha$ bits.
The memory peak is reached when we have to copy the data from the penultimate table to the final hash table with the above size.
The memory peak is at most $M(3+ \lg(m\alpha\sigma)) + M/2(2+\lg(m\alpha\sigma)) \le (2z-1)(11+3\lg(z \alpha \sigma))/2\alpha$.

If we compare this peak with the approach using a classic hash table (where we need to store the full key), we get a size of 
$M( \lg m + \lg m + \lg \sigma) + M/2 (\lg (m/2) + \lg(m/2) + \lg \sigma) \le 3 (2z-1) (4/3 + \lg(\sigma z^2))/\alpha $ bits.

This gives the following theorem:
\begin{theorem}
	We can compute the LZ78 and LZW factorization online using linear time with high probability and
	at most 
	$z(3\lg(z\sigma \alpha)+11)/\alpha$
	bits of working space, for a fixed~$\alpha$ with $0 < \alpha < 1$.
\end{theorem}

For the evaluation, we use a preliminary version of the implementation of~\citet{PoyiasPR17} that is based on~\cite{Cleary84} with the difference that \citeauthor{Cleary84} uses bidirectional probing (\cite{PoyiasPR17} uses linear probing).

\subsubsection{Rolling Hashing}
Here, we present an alternative trie representation with hashing, called \vRolling{}. %
The idea is to maintain the Karp-Rabin fingerprints~\cite{karp87efficient} of all computed factors in a hash table such that
the navigation in the trie is simulated by matching the fingerprint of a substring of the text with the fingerprints in the hash table.
Given that the fingerprint of the substring~$T[i..i+\ell-1]$ matches the fingerprint of a node~$u$,
we can compute the fingerprint of~$T[i..i+\ell]$ to find the child of~$u$ that is connected to~$u$ by an edge with label~$T[i+\ell]$.
To compute the fingerprints, we choose one of the two rolling hash function families:
\begin{itemize}
	\item a function of the randomized Karp-Rabin \vIDThree{} family~\cite{LemireK10}\footnote{\url{https://github.com/lemire/rollinghashcpp}}, and
	\item the function \vFermat{}$(T) = \sum_{i=1}^{\abs{T}} (T[i]-1) (\sigma+1)^{\abs{T}-i} \mod 2^w$, where
		the modulo by the word size~$w$ surrogates the integer overflow, and $T[i] - 1$ is in the range $[0..\sigma-1]$.
		In the case of a byte alphabet, $\sigma+1 = 2^8+1 = 257$ is a Fermat prime.
		We compute \vFermat{}$(T)$ with Horner's rule.
\end{itemize}
The LZ78/LZW computation using \vRolling{} is a Monte Carlo algorithm, since the computation can produce a wrong factorization if the computed fingerprints of two different strings are the same (because the fingerprints \emph{are} the hash table keys).

\subsubsection{Summary}
We summarize the description of the trie data structures in this and the previous section by \Cref{figSpace} 
showing the maximum space consumption of each described trie.
The maximum memory consumption is due to the peak at the last enlargement of the dynamic trie data structure, i.e., 
when the trie enlarges its space such that $z \le m \le 2z-1$ (where~$m$ is the number of elements it can maintain).

\begin{figure}[t]
	\floatbox[{\capbeside\thisfloatsetup{capbesideposition={right,top},capbesidewidth=3.2cm}}]{figure}[\FBwidth]%
	{%
	\caption{Upper and lower bound of the maximum memory used %
		during an LZ78/LZW factorization with~$z$ factors. 
		The size of a fingerprint is~$w$ bits.
	}
	\label{figSpace}
}{%
		\begin{tabular}{lll}
			\toprule
			Trie & Space Best Case (bits) & Space Worst Case (bits) %
			\\\midrule
\vBinary{} &
$3z(\lg(z^2\sigma) - 2/3)/2$ &
$3z(\lg(z^2\sigma) + 4/3)$ 
\\
\vTernary{} &
$3z (\lg(z^3\sigma) - 1)/2$ &
$3z (\lg(z^3\sigma) + 2)$ 
\\
\vHash{} &
$3z (\lg(z^2 \sigma) - 2/3)/2\alpha$ &
$6z (\lg(z^2 \sigma) + 4/3)/\alpha $ 
\\
\vCompact{} &
$3z(\lg(\alpha z\sigma)+8/3)/2\alpha$ &
$3z(\lg(\alpha z\sigma)+11/3)/\alpha$ 
\\
\vRolling{} &
$3z (w + \lg(z \sigma) - 1/3)/2\alpha$ &
$6z (w + \lg(z \sigma) + 2/3)/\alpha$ 
			\\\bottomrule
		\end{tabular}
		\hspace{-2em}
}%
\end{figure}

\section{Experiments and Conclusion} \label{sect:experiments}
We implemented the LZ tries in the tudocomp framework~\cite{tudocomp}\footnote{The source code of our implementations is freely available at~\url{https://github.com/tudocomp}, except for \vCompact{} and \vBonsai{} due to copyright restrictions.}.
The framework provides the implementation of an LZ78 and an LZW compressor.
Both compressors are parameterized by an LZ trie and a coder.
The coder is a function that takes the output of the factorization and generates the final binary output.
We selected the coder~{\tt bit} that stores the referred index~$y$ (with $y > 0$) of a factor~$F_x$ in $\upgauss{\lg x}$ bits.
That is because the factor~$F_x$ can have a referred index~$y$ only with~$y < x$.
We can restore the coded referred index on decompression since we know the index of the factor that we currently process and hence the number of bits used 
to store its referred index (if we coded it)\footnote{this approach is similar to \url{http://www.cplusplus.com/articles/iL18T05o}}.
This yields $\sum_i^z \upgauss{\lg i} = z\upgauss{\lg z} - (\lg e)z +\Oh{\lg z}$ bits for storing the (positive) referred indices.
The additional characters in LZ78 and the negative referred indices in LZW are output naively as 8-bit integers.

The LZ78 and LZW compressor are independent of the LZ trie implementation, i.e., all trie data structures described in the previous sections can be plugged into the LZW or LZ78 compressor easily.
We additionally incorporated the following trie data structures into tudocomp:
\begin{description}
	\item[\vCedar{}:] the Cedar trie~\cite{cedarTrie}, representing a trie using two arrays.
	\item[\vJudy{}:] the Judy array, advertised to be optimized for avoiding CPU cache misses (cf.~\cite{LuanDWNC07} for an evaluation). 
	\item[\vBonsai{}:] the m-Bonsai ($\gamma$) trie~\cite{mBonsai} representing a trie whose nodes are not labeled.
It uses a compact hash table, but unlike our approach, the key consists of the position of the parent in the hash table (instead of the label of the parent) and the character.
Due to this fact, we need to traverse the complete trie for enlarging the trie.
We store the labels of the trie nodes in an extra array.
\end{description}
The data structures are realized as \CPlusPlus{} classes.
We added a lightweight wrapper around each class providing the same interface for all tries.

\block{Online Feature}
Given an input stream with known length, we evaluate the online computation of the LZ78 and LZW compression for different LZ trie representations.
We assume that $\Sigma$ is a byte alphabet, i.e., $\sigma = 2^8$.
On computing a factor, we encode it and output it instantaneously.
This makes our compression program a filter~\cite{McIlroy1987RUR}, i.e., it processes the input stream and generates an output stream, 
buffering neither the input nor the output.

\block{Implementation Details}
The keys stored by \vHash{} are stored in integers with a width of $40$ bit.
The values stored by \vHash{}, \vRolling{} and \vBonsai{} are 32-bit integers.
For all variants working with hash tables, we initially set $\alpha$ to~$0.3$.

According to the birthday paradox, the likelihood that the fingerprints of two different substrings match is anti-correlated to the number of bits used for storing the fingerprint if we assume that the used rolling hash function distributes uniformly.
We used 64-bit fingerprints because, unlike 32-bit fingerprints, the factorization produced by \vRolling{} are correct for all test instances and all tested rolling hash functions.

\block{Hash Function}
We use \vCompact{} with a hash function of the LCG family.
Our hash table for \vHash{} uses a xorshift hash function\footnote{\label{footXorshift}\url{http://xorshift.di.unimi.it/splitmix64.c}} derived from~\cite{SteeleLF14}.
It is slower than simple multiplicative functions, but more resilient against clustering.
Alternatives are sophisticated hash functions like CLHash~\cite{LemireK16} or Zobrist hashing~\cite{zobrist,Lemire12}.
These are even more resilient against clustering, but have practical higher computation times in our experiments.%

\block{Setup}
The experiments were conducted on a machine with 32 GB of RAM, an
Intel Xeon CPU~\texttt{E3-1271~v3} and a Samsung SSD {\tt 850 EVO 250GB}.
The operating system was a 64-bit version of Ubuntu Linux~\num{14.04} with the kernel version~3.13.
We used a single execution thread for the experiments.
The source code was compiled using the GNU compiler {\tt g++~6.2.0} with the compile flags {\tt -O3 -march=native -DNDEBUG}.

\begin{figure}[t]
	\centering{%
\includegraphics[width=0.45\textwidth]{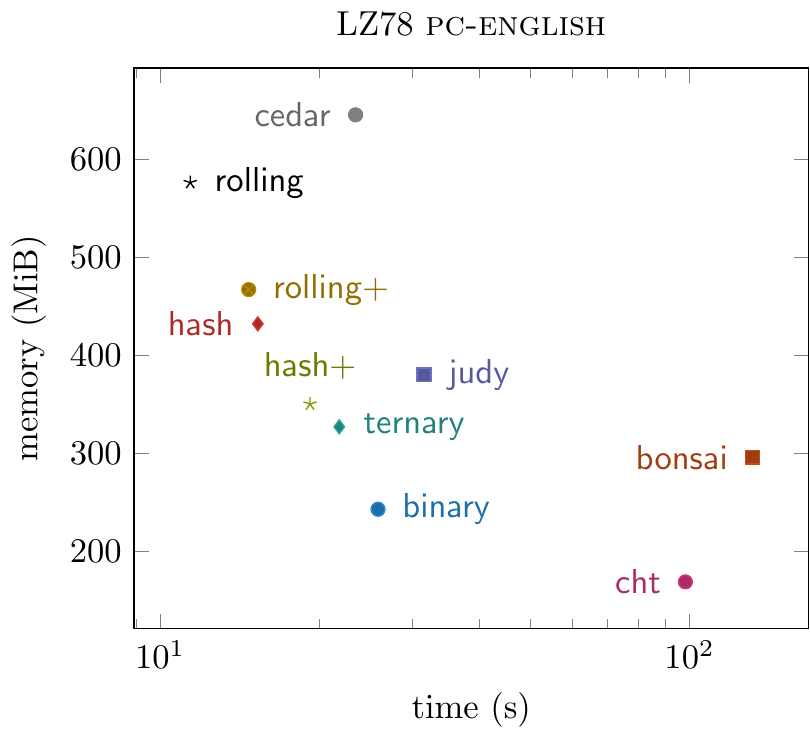}
\includegraphics[width=0.45\textwidth]{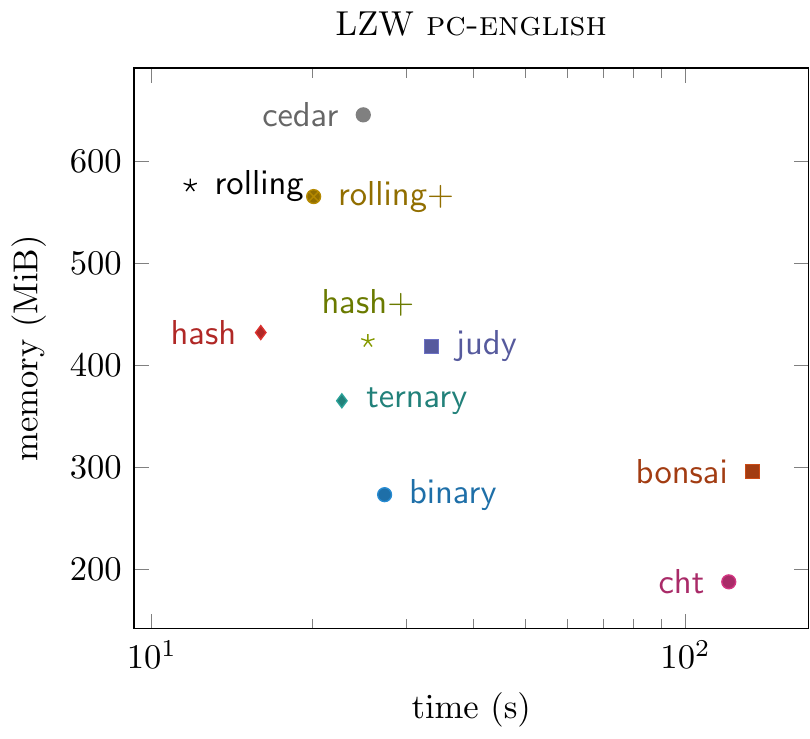}
	}
	\caption{Evaluation of LZ78/LZW on the English text \textsc{pc-english} with $\sigma = 226$.
	Left: LZ78 factorization with $z=21.4\textup{M}, \upgauss{\lg z} = 25$ and $z \lg z + z \lg \sigma = 83.8\textup{MiB}$. 
	The compressed file size is 80.2\textup{MiB}.
	Right: LZW factorization with $z=23.5\textup{M}, \upgauss{\lg (255+z)} = 25$ and $z \lg (255+z) = 70.13\textup{MiB}$.
	The compressed file size is 66.1\textup{MiB}.
	The LZ77 factorization consists of $z=14\textup{M}$ factors, and can be stored in $2z \lg n = 93.3\textup{MiB}$.
	}
	\label{figEnglish}
\end{figure}

\begin{figure}[t]
	\centering{%
\includegraphics[width=0.45\textwidth]{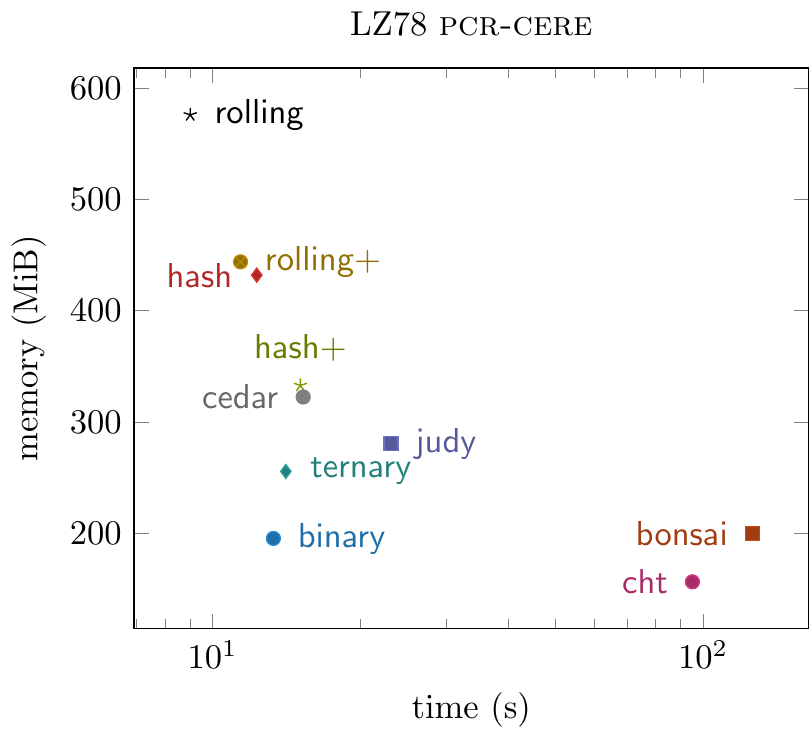}
\includegraphics[width=0.45\textwidth]{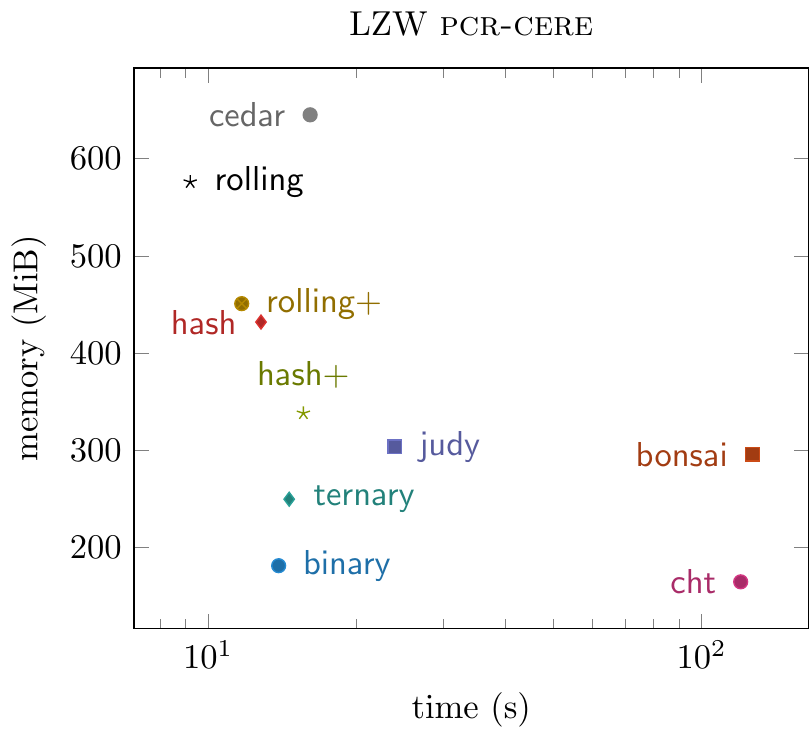}
	}
	\caption{Evaluation of LZ78/LZW on the repetitive DNA sequence \textsc{pcr-cere} with $\sigma = 6$.
	Left: LZ78 factorization with $z=15.8\textup{M}, \upgauss{\lg z} = 24$ and $z \lg z + z \lg \sigma = 50.0\textup{MiB}$. 
	The compressed file size is 80.2\textup{MiB}.
	Right: LZW factorization with $z=17.1\textup{M}, \upgauss{\lg (255+z)} = 25$ and $z \lg (255+z) = 50.9\textup{MiB}$.
	The compressed file size is 50.9\textup{MiB}.
	The LZ77 factorization consists of $z=1.38\textup{M}$ factors, and can be stored in $2z \lg n = 9.66\textup{MiB}$.
	}
	\label{figCere}
\end{figure}

\block{Datasets}
We evaluated the combinations of the aforementioned tries with the LZW and LZ78 algorithms on the 200MiB text collections provided by tudocomp.
We assume that the input alphabet is the byte alphabet ($\sigma = 2^8$).
The indices of the factors are represented with 32-bit integers.
Due to space restrictions, we choose \textsc{pc-english} as a representative for a standard English text and \textsc{pcr-cere} as a representative for a highly-repetitive text with small alphabet size (the evaluation on the other datasets were quite similar).
We plotted the memory consumption against the time (in logarithmic scale) for both datasets in \Cref{figEnglish} and \Cref{figCere}.
To avoid clutter, we selected one hash function per rolling hash table:
We chose \vFermat{} with \vRolling{} and \vIDThree{} with \vRolling{}+ for the plots.
We additionally added the number of LZ77 factors~\cite{ziv77universal} as a reference.

\block{Overall Evaluation}
The evaluation shows that the fastest option is \vRolling{}. 
The size of its fingerprints is a trade-off between space and the probability of a correct output.
When space is an issue, \vRolling{} with 64-bit fingerprints is no match for more space saving trie data structures.
\vHash{} is the second fastest LZ trie in the experiments.
With 40-bit keys it uses less memory than \vRolling{}, but is slightly slower.
Depending on the quality of the resize hint, the variants \vHash{}+ and \vRolling{}+ take 50\% up to 100\% of the size of \vHash{} and \vRolling{}, respectively.
\vHash{}+ and \vRolling{}+ are always slower than their respective standard variants, sometimes
slower than the deterministic data structures \vTernary{} and \vBinary{}.
\vBinary{}'s speed excels at texts with very small alphabets, while \vTernary{} usually outperforms \vBinary{}.
Only \vCompact{} can compete with \vBinary{} in terms of space, but is magnitudes slower than most alternatives.
The third party data structures~\vCedar{}, \vBonsai{} and \vJudy{} could not make it to the Pareto front.

\block{Evaluation of \vRolling{}}
The hash table with the rolling hash function \vFermat{} is slightly faster than with a function of the \vIDThree{} family, 
but the hash table with \vFermat{} tends to have more collisions (cf. \Cref{tableCollisions}).
It is magnitudes slower at less compressible texts like \textsc{pc-proteins} due to the high occurrence of collisions.
The number of collisions can drop if we post-process the output of \vFermat{} with a hash function that is more collision resistant
(like the hash function used for \vHash{}).
Applying the hash function on \vFermat{} speeds up the computation only if the number of collisions is sufficiently high (e.g., \vRolling{}+ with \vFermat{} in \Cref{tableCollisions}).

The domain of the Karp-Rabin fingerprints can be made large enough to be robust against collisions when hashing large texts.
In our case, 64-bit fingerprints fitting in one computer word were sufficient.
Checking whether the factorization is correct can be done by reconstructing the text with the output and the built LZ trie.
However, a compression with \vRolling{} combined with a decompression step takes more time than other approaches like \vHash{} or \vBinary{}.
Hence, a Las Vegas algorithm based on \vRolling{} is practically not interesting.

\begin{table}[t]
	\floatbox[{\capbeside\thisfloatsetup{capbesideposition={right,center},capbesidewidth=4cm}}]{table}[\FBwidth]%
	{%
	\caption{Detailed evaluation of the tries using hashing. 
We evaluated the number of collisions and the final table size~$M$ for the LZ78 factorization of 200 MiB \textsc{pc-english}. 
\xor{} denotes that the output of the rolling hash function is plugged into the xorshift hash function used by \vHash{}.
An entry in \vRolling{} costs $64+32$ bits, an entry in \vHash{} $40+32$ bits.
}
	\label{tableCollisions}
}{%
\renewcommand{\tabcolsep}{0.20cm}
		\begin{tabular}{lrrrr}
			\toprule
Trie             & \#Collisions & $M$      & memory & time
\\\midrule
\multicolumn{4}{l}{\vRolling{} with} \\
- \vIDThree{}      & 36M          & 33.6M & 576.0MiB & 11.6s \\
- \vFermat{}       & 137M         & 33.6M & 576.0MiB & 11.4s \\
- \vFermat{}\xor{}  & 36M          & 33.6M & 576.0MiB & 11.8s \\
\multicolumn{4}{l}{\vRolling{}+ with} \\
- \vIDThree{}     & 140M         & 24.0M & 466.9MiB & 14.7s \\
- \vFermat{}       & 938M         & 24.0M & 466.9MiB & 21.0s \\
- \vFermat{}\xor{} & 142M         & 24.0M & 466.9MiB & 15.8s \\
\midrule
\vHash{}         & 36M          & 33.6M & 432.0MiB & 15.3s \\
\vHash{}+        & 137M         & 24.0M & 350.2MiB & 19.1s
			\\\bottomrule
		\end{tabular}
		\hspace{-2em}
}%
\end{table}

\block{Outlook}
An interesting option is to switch from the linear probing scheme to a more sophisticated scheme 
whose running time is stable for high loads, too~\cite{2017arXiv170500997M}.
This could be especially beneficent if the resize hint provides a more accurate lower bound on the number of factors.

Speaking of novel hash tables, we could combine the compact hash table~\cite{Cleary84} with the
memory management of Google's sparse hash table\footnote{\url{https://github.com/sparsehash/sparsehash}} 
leading to an even more memory friendly trie representation.

\JO{%
\block{Outlook}
There is much space for extending the presented evaluation.
For instance, we could squeeze the values stored by \vHash{} and \vRolling{} in $\upgauss{\lg z}$ bits instead of storing them in 32-bit integers, 
sacrificing speed for memory.
Another way would be to change the experimental model.
Instead of parsing the text in one shot, we could perform a precomputation step in which we determine
the number and the frequency of the characters in the text. This information can be exploited to store the characters more succinctly.

Speaking of novel hash tables, we could combine the compact hash table~\cite{Cleary84} with the
memory management of Google's sparse hash table\footnote{\url{https://github.com/sparsehash/sparsehash}} 
leading to an even more memory friendly trie representation.

Overall, the experimental results are not satisfying as 
theoretical results advertise dynamic trie data structures working in \Oh{z \lg z} bits 
while still allowing us to compute the LZ78/LZW factorization in nearly linear time.
}

\block{Acknowledgements}
We are grateful to Patrick Dinklage for spell-checking the paper, and to Marvin Löbel for providing 
the basement of the LZ78/LZW framework in tudocomp.
Further, we thank Andreas Poyias for sharing the source code of the m-Bonsai trie~\cite{mBonsai} and the compact hash table~\cite{PoyiasPR17}.

\putbib
\end{bibunit}
\newpage
\begin{bibunit}
\appendix

\begin{table}[t]
	\centerline{%
		\begin{tabular}{l*{8}{r}}
			\toprule
			&	\multicolumn{4}{c}{\textsc{pc-english}} & \multicolumn{4}{c}{\textsc{pcr-cere}}
			\\ \cmidrule(lr){2-5} \cmidrule(lr){6-9}
			&  \multicolumn{2}{c}{LZ78} &  \multicolumn{2}{c}{LZW} & \multicolumn{2}{c}{LZ78} &  \multicolumn{2}{c}{LZW} 
			\\ \cmidrule(lr){2-3} \cmidrule(lr){4-5} \cmidrule(lr){6-7} \cmidrule(lr){8-9}
Trie  & Time & Space & Time & Space & Time & Space  & Time & Space 
			\\\midrule
\multicolumn{4}{l}{\vHash{} with hash table}
\\
			{\tt std::unordered\_map} & 51.0s & 856.6MiB & 54.0s & 937.9MiB & 42.3s & 703.2MiB & 44.1s & 760.8MiB 
			\\
			{\tt std::map} & 161.2s & 980.2MiB & 167.2s & 1.1GiB & 98.8s & 722.5MiB & 104.6s & 781.6MiB
			\\
			{\tt rigtorp}\footnote{\url{https://github.com/rigtorp/HashMap}, $\alpha = 0.5$ hard coded} & 14.9s & 960.0MiB  & 15.2s & 960.0MiB & 12.0s & 960.0MiB & 12.3s & 960.0MiB
			\\
			{\tt flathash}\footnote{\label{footFlathash}\url{https://probablydance.com/2017/02/26/i-wrote-the-fastest-hashtable/}, it uses the identity as a hash function and doubles its size when experiencing too much collisions} & 33.5s & 24GiB & 24.5s & 24GiB & 18.5s & 6GiB & 19.2s & 6GiB
\\
			{\tt flathash}\footnote{See \Cref{footFlathash}, but with our default hash function} & 15.1s & 1.3GiB & 15.7s & 1.3GiB  & 12.4s & 1.3GiB  & 13.0s & 1.3GiB
\\
{\tt densehash}\footnote{\label{footSparsehash}\url{https://github.com/sparsehash/sparsehash}} & 23.0s & 576.0MiB & 24.4s & 576.0MiB & 29.4s & 576.0MiB & 30.8s & 576.0MiB\\
{\tt sparsehash}$^{\text{\ref{footSparsehash}}}$ & 49.1s & 255.7MiB & 52.2s & 280.0MiB & 68.6s & 191.3MiB & 72.4s& 206.1MiB
\\\midrule
			LZ-index~\cite{Navarro08}  & 24.6s & 1047MiB & & & 14.5s & 817.3MiB  
			\\\bottomrule
		\end{tabular}
	}%
\caption{\vHash{} with different hash tables, and the LZ-index. }
	\label{tableAlternatives}
\end{table}
\section{Variations of Hash Tables}
The trie representation \vHash{} can be generalized to be used with any associative container.
The easiest implementation is to use the balanced binary tree {\tt std::map} or the hash table {\tt std::unordered\_map} provided by the standard library of \CPlusPlus{11}.
{\tt std::unordered\_map} is conform to the interface of the \CPlusPlus{} standard library, but therefore sacrifices performance.
It uses separate chaining that tends to use a lot of small memory allocations affecting the overall running time (see \Cref{tableAlternatives}).
Another pitfall is to use the standard \CPlusPlus{11} hash function for integers that is just the identity function.
Although this is the fastest available hash function,
it performs poorly in the experiments.
There are two reasons.
The first is that $k \mapsto k \mod M$ badly distributes the tuples if~$M$ is not a prime.
The second is that the input data is not independent:
In the case of LZ78 and LZW, the composed key~$c + \ell\sigma$ of a node~$v$ connected to its parent with label~$\ell$ by an edge with label~$c$
holds information about the trie topology: all nodes whose keys are $\ell\sigma+d$ for a~$d\in\Sigma$ are the siblings of~$v$. 
Since~$\ell$ is smaller than the label of~$v$ ($\ell$ is the referred index of the factor corresponding to~$v$),
larger keys depend on the existence of some keys with smaller values.
Both problems can be tackled by using a hash function with an avalanche effect property, i.e.,
flipping a single bit of the input changes roughly half of the bits of the output.
In \Cref{tableAlternatives} we evaluated the identity and our standard hash function (see \Cref{footXorshift}) as hash functions for the hash table~{\tt flathash},
which seems to be very sensitive for hash collisions.

We selected the LZ trie of the LZ-index~\cite{Navarro08} as an external competitor in \Cref{tableAlternatives}.
We terminated the execution after producing the LZ trie of the LZ78 factorization.  
We did not integrate this data structure into tudocomp.

The only interesting configuration is \vHash{} with the hash table {\tt sparsehash}, 
since it takes 4.1MB less space than \vBinary{} while still being faster than \vCompact{}, at the LZ78-factorization of \textsc{pcr-cere}.

\begin{figure}[t]
	\centering{%
\includegraphics[width=0.45\textwidth]{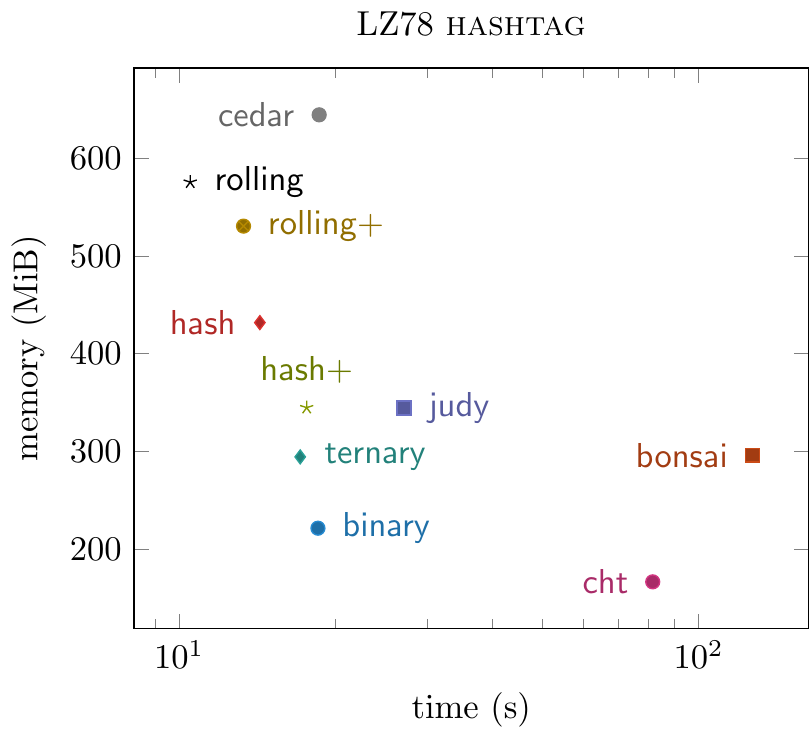}
\includegraphics[width=0.45\textwidth]{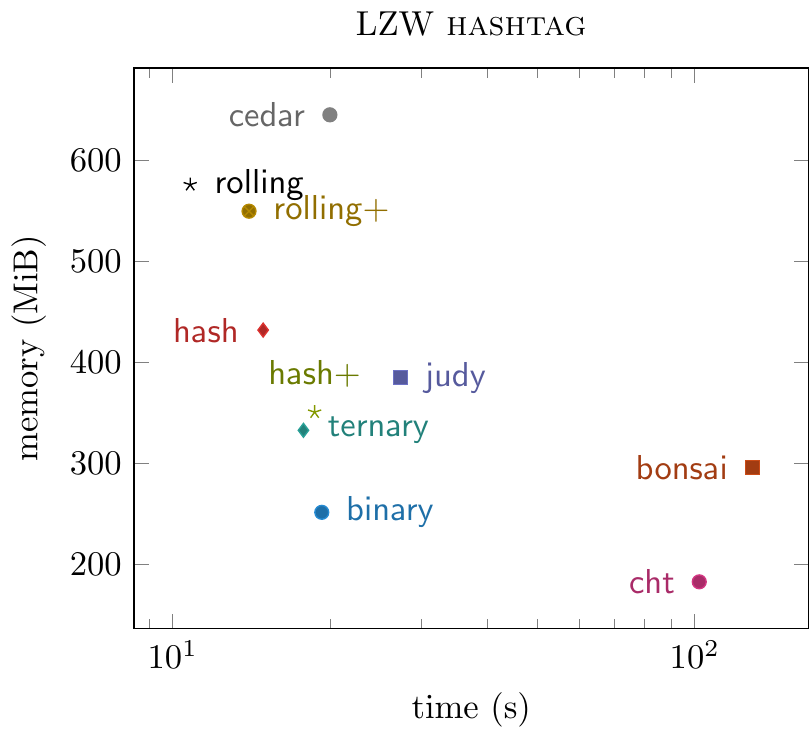}
	}
	\caption{%
	Evaluation of LZ78/LZW on the tab-spaced-version file \textsc{hashtag} with $\sigma = 179$.
	Left: LZ78 factorization with $z=18.9\textup{M}, \upgauss{\lg z} = 25$ and $z \lg z + z \lg \sigma = 73.4\textup{MiB}$. 
	The compressed file size is 70.6\textup{MiB}.
	Right: LZW factorization with $z=21.1\textup{M}, \upgauss{\lg (255+z)} = 25$ and $z \lg (255+z) = 62.9\textup{MiB}$.
	The compressed file size is 58.9\textup{MiB}.
	The LZ77 factorization consists of $z=13.7\textup{M}$ factors, and can be stored in $2z \lg n = 90.4\textup{MiB}$.
	}
	\label{figHashtag}
\end{figure}

\begin{figure}[t]
	\centering{%
		\includegraphics[width=0.45\textwidth]{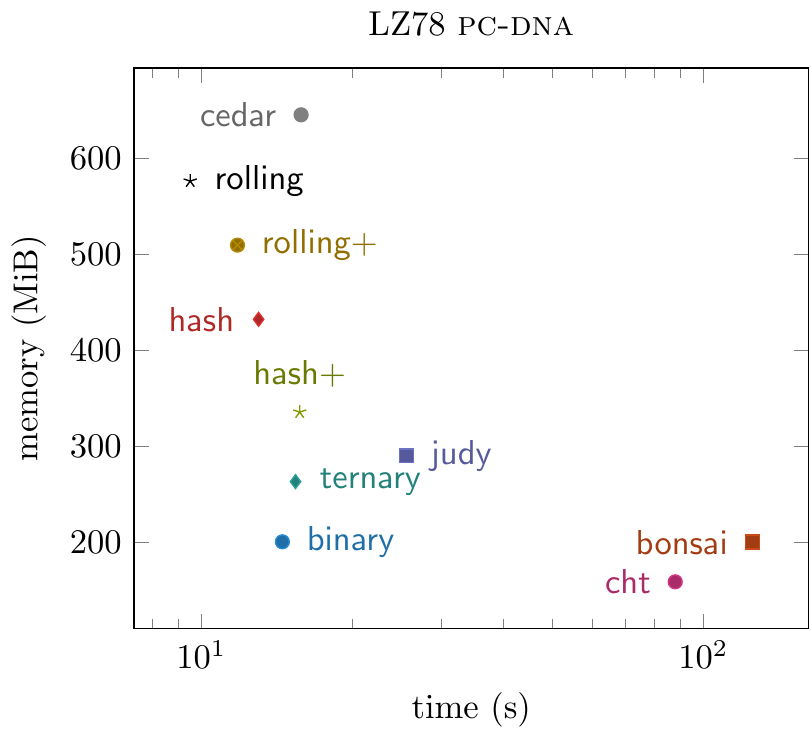}
		\includegraphics[width=0.45\textwidth]{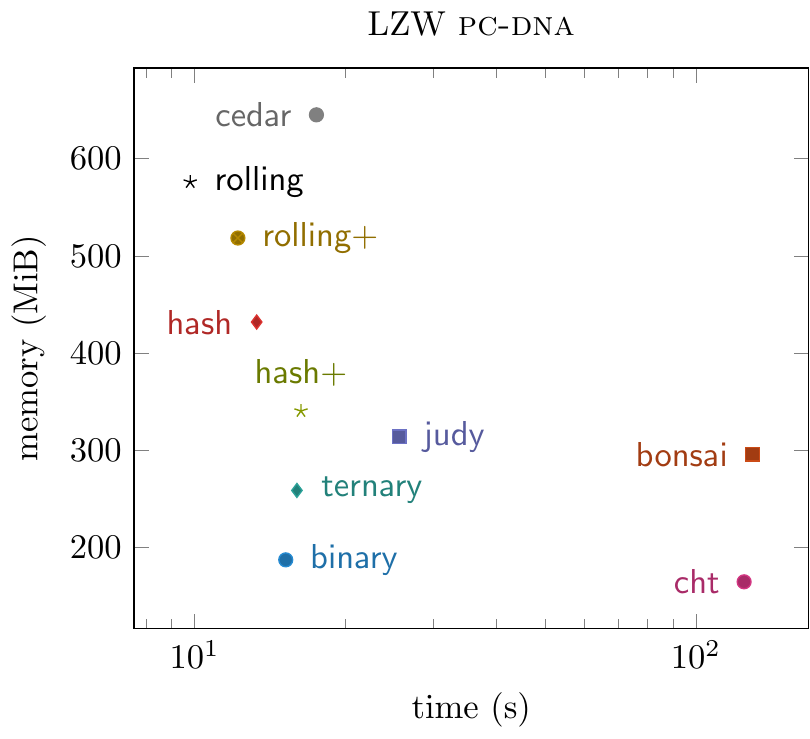}
	}
	\caption{%
	Evaluation of LZ78/LZW on the DNA sequence \textsc{pc-dna} with $\sigma = 17$.
	Left: LZ78 factorization with $z=16.4\textup{M}, \upgauss{\lg z} = 24$ and $z \lg z + z \lg \sigma = 54.8\textup{MiB}$. 
	The compressed file size is 60.5\textup{MiB}.
	Right: LZW factorization with $z=17.8\textup{M}, \upgauss{\lg (255+z)} = 25$ and $z \lg (255+z) = 52.9\textup{MiB}$.
	The compressed file size is 48.9\textup{MiB}.
	The LZ77 factorization consists of $z=13.9\textup{M}$ factors, and can be stored in $2z \lg n = 92.1\textup{MiB}$.
	}
	\label{figDna}
\end{figure}

\section{Reasons for Linear Probing}
Collisions in our hash table are resolved by linear probing.
Linear probing inserts a tuple with key~$k$ at the first free entry, starting with the entry at index~$h(k) \mod M$.
Linear probing is cache-efficient if the keys have a small bit width (like fitting in a computer word).
Using large hash tables and small keys, the cache-efficiency can compensate the chance of higher collisions~\cite{Askitis09,HeilemanL05}.
Linear probing excels if the load ratio is below 50\%, and it is still competitive up to a load ratio of 80\%~\cite{2017arXiv170500997M,BlackMQ98}.
Nevertheless, its main drawback is \emph{clustering}:
Linear probing creates runs, i.e., entries whose hash values are equal. With a sufficient high load, it is likely that
runs can merge such that long sequence of entries with different hash values emerge.
When trying to look up a key~$k$, we have to search the sequence of succeeding elements starting at the initial address until finding a tuple whose key is~$k$, or ending at an empty entry.
Fortunately, the expected time of a search is rather promising for an $\alpha$ not too close to one: 
Given that the used hash function~$h$ distributes the keys independently and uniformly, we get \Oh{1/(1-\alpha)^2} expected time for a search~\cite{knuthIII}.
In practice, even weak hash functions (like we use in this article) tend to behave as truly independent hash functions~\cite{ChungMV13}.
These properties convinced us that linear probing is a good candidate for our representations of the LZ trie using a hash table.

\section{More Evaluation}
We additionally evaluated the presented trie data structures on two other datasets in \Cref{figHashtag,figDna},
showing similar characteristics as the plots in \Cref{figCere,figEnglish}.

\putbib
\end{bibunit}
\end{document}